\documentclass[aps,prb,twocolumn,showpacs]{revtex4-1}
\usepackage{graphicx}
\usepackage{dcolumn}
\usepackage{bm}
\usepackage{amsmath}
\usepackage{multirow}
\usepackage{tabularx}
\usepackage{rotating}  % used by me
\usepackage{hyperref}  % used by me
\usepackage{braket}
\usepackage{xcolor}
\usepackage{soul}
\begin{document}
\author{Debashish Das}\email{das.debashish37@gmail.com}
\affiliation{Department of Physics, Indian Institute of Technology, Bombay, Powai, Mumbai 400076, India}
\author{Aftab Alam}\email{aftab@iitb.ac.in}
\affiliation{Department of Physics, Indian Institute of Technology, Bombay, Powai, Mumbai 400076, India}

\title{Exotic multiferroic properties of spinel structured $AB_2O_4$ compounds: A Monte Carlo Study}

\begin{abstract}
Spinel structured compounds, $AB_2O_4$, are special because of their exotic multiferroic properties. In $ACr_2O_4$ ($A$=$Co$, $Mn$,$Fe$), a switchable polarization has been observed experimentally due to a non-collinear magnetic spin order. In this article, we demonstrated the microscopic origin behind such magnetic spin order, hysteresis, polarisation and the  so-called magnetic compensation effect in $ACr_2O_4$ ($A$=$Co$, $Mn$,$Fe$, $Ni$) using Monte Carlo simulation. With a careful choice of the exchange interaction, we were able to explain various experimental findings such as magnetization vs. temperature (T) behavior, conical stability, unique magnetic ordering and polarization in a representative compound $CoCr_2O_4$ which is the best known multiferroic compound in the  $AB_2O_4$ spinel family. We have also studied the effect of $Fe$-substitution in $CoCr_2O_4$, with an onset of few exotic phenomena such as magnetic compensation and sign reversible exchange bias effect. These effects are investigated using an effective  interactions mimicking the effect of substitution. Two other compounds in this family, $CoMn_2O_4$ and $CoFe_2O_4$, are also studied where no conical magnetic order and polarisation was observed, as hence provide a distinct contrast. Here all calculations are done using the polarisation calculated by the spin-current model. This model has certain limitation and it works quite good for low temperature and low magnetic field. But the model despite  its limitation it can reproduce sign reversible exchange bias and magnetic compensation like phenomena quite well.         
\end{abstract}
\maketitle

\section{Introduction}
$CoCr_2O_4$ is a classic example of spinel which is observed to show a new kind of polarisation at very low temperature, whose origin lies in the formation of a conical magnetic order.\cite{Pol-org} The application of the magnetic field manipulates the cone angle and hence the coupling between ferromagnetism and ferroelectric properties. Similar multiferroism has been reported for other spinel compounds such as $MnCr_2O_4$,\cite{Tomiyasu} $NiCr_2O_4$,\cite{ACr2O4-pol} and $FeCr_2O_4$.\cite{FeCr2O4-pol} These four spinels posses both polarisation and magnetism due to spin origin. However, there are several other compounds, $RMnO_3$ (R= Tb, Dy), in perovskite family where the polarisation is due to spin spiral developed in the plane.\cite{RMnO3_1, RMnO3_2} Therefore, such a compound does not have any net magnetization (M). However, the conical magnetic order in $ACr_2O_4$ adds an extra magnetism along the cone axis and makes these compounds much more interesting.

There have been some experiments on this class of  $AB_2O_4$ compounds, which provide useful information about their novel properties. Yamasaki \textit{et al.}\cite{Pol-org} reported the signature of polarisation in $CoCr_2O_4$ below $T_s$=27 K. They also showed how polarisation can be controlled using magnetic field. Neutron scattering experiments on $ACr_2O_4$ [$A$=$Co$, $Mn$] was first performed by Tomiyasu \textit{et al.},\cite{Tomiyasu} who estimated the cone angle by analyzing the experimental intensity of satellite reflections. They also proposed a unique concept of ``Weak Magnetic Geometrical Frustration" (MGF) in spinel $AB_2O_4$, where both $A$ and $B$ cation are magnetic. Such weak MGF is responsible for the short-range conical spiral. Using neutron diffraction, Chang \textit{et al.}\cite{incommensurate} predicted a transformation from incommensurate conical spin order to commensurate order in $CoCr_2O_4$ at lowest temperature. A complete understanding of such transformation is lacking in the literature. Spin current model\cite{spin-current} is one simplistic approach which provides some conceptual advancement about incommensurate conical spin order, however, a firm understanding of incommensurate to commensurate transformation requires a better model.

These class of compounds show few other phenomena such as negative magnetization, magnetic compensation and sign reversible exchange bias at a critical temperature called magnetic compensation temperature ($T_{comp}$).\cite{padam-Fe,ram-Mn, Junmoni-Mn-Fe, Junmoni-Ni-Al, Junmoni-Ni-Fe1, Junmoni-Ni-Fe2} This is a temperature at which different sublattice magnetization cancels each other to fully compensate the net magnetization (M=0). Interestingly, it changes sign if one goes beyond this temperature. Depending on the substituting element, in some cases, magnetic compensation is associated with the exchange bias phenomena. Such unique phenomena are very useful for magnetic storage devices which require a reference fixed magnetization direction in space for switching magnetic field. Compounds having exchange bias are highly suitable for such a device because their hysteresis is not centred at M=0, H=0, rather shifted towards +ve or -ve side. Although the phenomena of exchange bias are well understood in various compounds including FM/AFM layered compounds,\cite{EB} the same is not true for the substituted spinel compounds which crystallize in a single phase. A deeper understanding of all these exotic phenomena is highly desired.

Using the generalized Luttinger-Tisza\cite{GLT} method,  a conical ground state can be found theoretically,\cite{LKDM} by defining a parameter $u$

\[
u=\frac{4J_{BB}S_B}{3J_{AB}S_A}
\]

Here $S_A$ and $S_B$ are the A-site (tetrahedral) and B-site (Octahedral) magnetic spins, $J_{AB}$ and $J_{BB}$ represent the exchange interaction between first nearest neighbor $A$-$B$ and  $B$-$B$ pairs respectively. According to the theory, the stable conical spin order is possible only if $u$ lies between $0.88$ and $1.298$.

Yan \textit{et al}\cite{Yao-2009,Yao-2009-2,Yao-2010,Yao-2011,Yao-2013,Yao-2017} has studied the conical spin order by performing simulation on a 3-dimensional spinel lattice. They show that $\hat{J}_{BB}$  and $\hat{J}_{AA}$ enhance the spin frustration, and single ion anisotropy helps to stabilize the cone state. Here $\hat{J}_{ij}=J_{ij}|\overrightarrow{S_i}|.|\overrightarrow{S_j}|$ and is called magnetic coupling constant.

In this article, the conical spin order of $ACr_2O_4$ ($A$=$Mn$, $Fe$, $Co$ and $Ni$) along with  $CoMn_2O_4$ and $CoFe_2O_4$ are studied  using a combined Density Functional Theory (DFT) and Monte Carlo based Metropolis algorithm. The latter two compounds do not show conical spin order.  For these six compounds, we have calculated the exchange interactions using the self consistent Density Functional Theory. We have then varied the interaction parameters and found a new set of exchange interactions which best fit the experimental magnetization and hysteresis curves. For comparison sake, the investigation of magnetic ordering, magnetization, hysteresis curve, and the ground state spin order were carried out using both sets of exchange interactions. We have also simulated the magnetic compensation and exchange bias behavior around $T_{comp}$. We found an effective exchange interaction pairs for the system $CoCr_2O_4$, for which its magnetization is similar to $Fe$ substituted $CoCr_2O_4$ showing magnetic compensation effect followed by a turn over in the sign having of M. Using these sets of exchange bias, we are able to predict the sign reversible exchange bias at around $T_{comp}$, as observed experimentally.\cite{padam-Fe}

\begin{table*}
    \caption{\label{table1} For six $AB_2O_4$ spinel compounds; coupling constants ($\hat{J}_{BB}$, $\hat{J}_{AB}$ and $\hat{J}_{AA}$), conical spin order parameter ($u$), magnetic moments at A and B sites calculated from DFT and Montecarlo simulations, transition temperature ($T_c$) obtained from simulation and experiments.  Each of these properties are calculated with two sets of coupling constants (set-1 and set-2), as described in the text. For $CoMn_2O_4$, the $Mn$-$Mn$ bonds in the xy-plane are smaller compared to the $Mn$-$Mn$ bond out of plane, therefore I and O represents in-plane and out-of plane $\hat{J}_{BB}$ interaction. For $CoFe_2O_4$, which crystallizes in inverse spinel structure, half of $B$-site are filled with $Co$ and other half be $Fe$, A sites are completely filled by $Fe$. This creates three types of $B$-$B$ interactions ($Co$-$Co$, $Fe$-$Fe$, \& $Co$-$Fe$) and 2 types of $A$-$B$ interactions ($Fe$-$Co$, \& $Fe$-$Fe$).
    }
 
\begin{center}
        \begin{tabular}{|c|c|c|c|c|c|c|c|c|c|c|c|} \hline \hline
                System    &   &\multicolumn{4}{c|}{ Coupling constant} &  \multicolumn{2}{c|}{Moment} &  \multicolumn{2}{c|}{Moment}& Calculated & Expt. \\ 
                          &   &\multicolumn{4}{c|}{}            & \multicolumn{2}{c|}{(DFT)} &  \multicolumn{2}{c|}{ (Monte Carlo)} &            &  \\   \hline                        
                          &   &$\hat{J}_{BB}$ &$\hat{J}_{AB}$ & $\hat{J}_{AA}$  & $u$ & $M_A$ & $M_B$  & $M_A$ & $M_B$              & $T_c$      & $T_c$  \\ 
                          &   &     (meV)     & (meV)         &   (meV)         &     &($\mu_B$)&($\mu_B$)&($\mu_B$)&($\mu_B$)              & (K)       & (K) \\ \hline \hline               
               $MnCr_2O_4$& set 1&  -1.74  & -1.28   & -1.58  &1.81 & -4.50 & 3.01   & -5    & 3  & 40 & \multirow{ 2}{*}{51\cite{Tomiyasu}} \\ 
                          & set 2&  -0.97  & -0.85   &  0.00  &1.52 &       &        &       &    & 42 &\\ \hline
               $FeCr_2O_4$& set 1&  -2.88  & -2.83   & -0.67  &1.35 & -3.69 & 2.95   &-4    & 3   & 117& \multirow{ 2}{*}{74\cite{FeCr2O4-Tc}}\\ 
                          & set 2&  -1.38  & -1.94   & -0.67  &0.95 &       &        &      &     & 103& \\  \hline
               $CoCr_2O_4$& set 1&  -3.01  & -3.26   & -0.56  &1.23 & -2.60 & 3.04   &-3    & 3   & 145& \multirow{ 2}{*}{97\cite{CoCr2O4-Tc}}\\ 
                          & set 2& -4.25   &-2.83    & 0.00   &2.00 &       &        &     &      & 94 &\\ \hline
               $NiCr_2O_4$& set 1&  -5.36  & -3.94   & -1.64  &1.81 & -1.69 & 2.93   &-2    & 3   & 24 & \multirow{ 2}{*}{80\cite{NiCr2O4-Tc}}\\
                          & set 2&  -3.75  & -2.38   & 0.00   &2.10 &       &        &      &     & 80 & \\ \hline     
               $CoMn_2O_4$&set 1 & -9.46 (I) &-3.53 & -0.29 & 3.57 & -2.68 & 3.81&-3    & 4       & 52 & \multirow{ 2}{*}{85\cite{CoMn2O4-Tc}}\\
                          &      & -1.05 (O) &      &       & 0.40 &       &     &      &         &    & \\ 
                          &set 2 & -5.46 (I) &-3.53 & -0.29 &2.06  & -2.68 & 3.81&-3    & 4       & 60 & \\
                          &      & -3.05 (O) &      &       &1.15  &       &      &      &        &    & \\ \hline
               $CoFe_2O_4$&set 1 &  0.08 (Co-Co) & -10.43 (Fe-Co) &        &      & -3.98& 2.66, 4.10 &-4    & 3, 4  & 870 & \multirow{ 2}{*}{860\cite{CoFe2O4-Tc}}\\
                          &      & -4.77 (Fe-Fe) & -21.65 (Fe-Fe) &   -2.06 &0.29 &      &     &      &              &     & \\
                          &      &  0.84 (FeCo)  &                &         &     &      &     &      &              &     & \\ 
                          &set 2 &  0.08 (Co-Co) & -10.00 (Fe-Co) &         &     &      &     &-4    & 3, 4         & 840 &  \\
                          &      & -4.77 (Fe-Fe) & -10.00 (Fe-Fe) & -2.06   &0.63 &      &     &      &              &     &  \\
                          &      &  0.84 (FeCo)  &                &         &     &      &     &      &              &     & \\ \hline \hline
            \end{tabular}
        \end{center}
\end{table*}

\section{Methodology}
For calculation, we have generated a 3-dimensional spinel structure involving a  7$\times$7$\times$7 supercell of 2 formula unit which contains a total of 2058 numbers of magnetic atoms. Oxygen atoms are removed while generating the supercell as they don't  contribute to magnetisation. We defined the energy equation of the form 
\begin{equation}
E=-\sum_{<i,j>} J_{ij} \overrightarrow{S_i}.\overrightarrow{S_j}-\overrightarrow{M}.\overrightarrow{h_m}-\overrightarrow{P}.\overrightarrow{h_e}
\end{equation} 

where $\overrightarrow{P}$ and $\overrightarrow{M}$ are polarisation and magnetisation respectively, defined as
\begin{equation}
\overrightarrow{P}=a.\sum_{<i,j>}\overrightarrow{e_{ij}}\times\overrightarrow{S_i}\times\overrightarrow{S_j}
\end{equation}

and
\begin{equation} 
\overrightarrow{M}=\sum_i \left( \sqrt{(S_i^x)^2+(S_i^y)^2+(S_i^z)^2}\right).g.\overrightarrow{\mu_B}
\end{equation}
where $\overrightarrow{e}_{ij}$ is the vector connecting $\overrightarrow{S}_i$ and $\overrightarrow{S}_j$, '$a$' is a proportionality constant and $g$ is the Land{\'e} $g$-factor which is 2 $\mu_B$. We solve this energy equation by Monte Carlo simulation where the spins are considered classical vectors that are updated by Metropolis algorithm. 1,00,000 steps are taken for equilibration and the average of last 5000 steps data are used to calculate physical quantities. $\sum_{<i,j>}$ is summation over nearest $B$-$B$, $A$-$B$ and $A$-$A$ type of neighbors, while the higher-order neighbors are neglected. For the calculation of temperature dependence of magnetization, we have taken 5000 Monte Carlo steps for each temperature and the temperature is increased in the steps of 1 K. To reach the correct conical ground state, we have applied a large electric field (~ 20000 kV/m along [110] directions) and a magnetic field (20 Tesla along [001] direction), as also used by Nehme \textit{et al}.\cite{Nehme}

\section{Result and Discussion}

 \subsection{Exchange Interaction parameters \& Magnetisation}
In order to simulate various system properties, we have calculated two sets of exchange interaction parameters. \\
(a) set-1: Interaction parameters derived from self-consistent first principles-based DFT calculation.\\
(b) set-2: A new set of interaction parameters which best fit the experimental magnetisation.\cite{padam-Fe}\\

Table \ref{table1} shows the above two sets of interaction parameters for six representative systems $ACr_2O_4$ ($A$= $Mn$, $Fe$, $Co$, and $Ni$), $CoMn_2O_4$ and $CoFe_2O_4$. For $CoMn_2O_4$, the $Mn$-$Mn$ bonds in the $xy$-plane are smaller compared to those which are out of plane. I and O represent in-plane and out-of-plane $\hat{J}_{BB}$ interactions for CoMn$_2$O$_4$. For $CoFe_2O_4$ which crystallized in inverse spinel structure, half of the $B$-site are filled with $Co$ and the rest by $Fe$. This geometry creates three types of $B$-$B$ interactions ($Co$-$Co$, $Fe$-$Fe$ \& $Co$-$Fe$) and two types of $A$-$B$ interactions ($Fe$-$Co$ \& $Fe$-$Fe$).    
 
\begin{center}
\begin{figure}[b]
\includegraphics[scale=0.4]{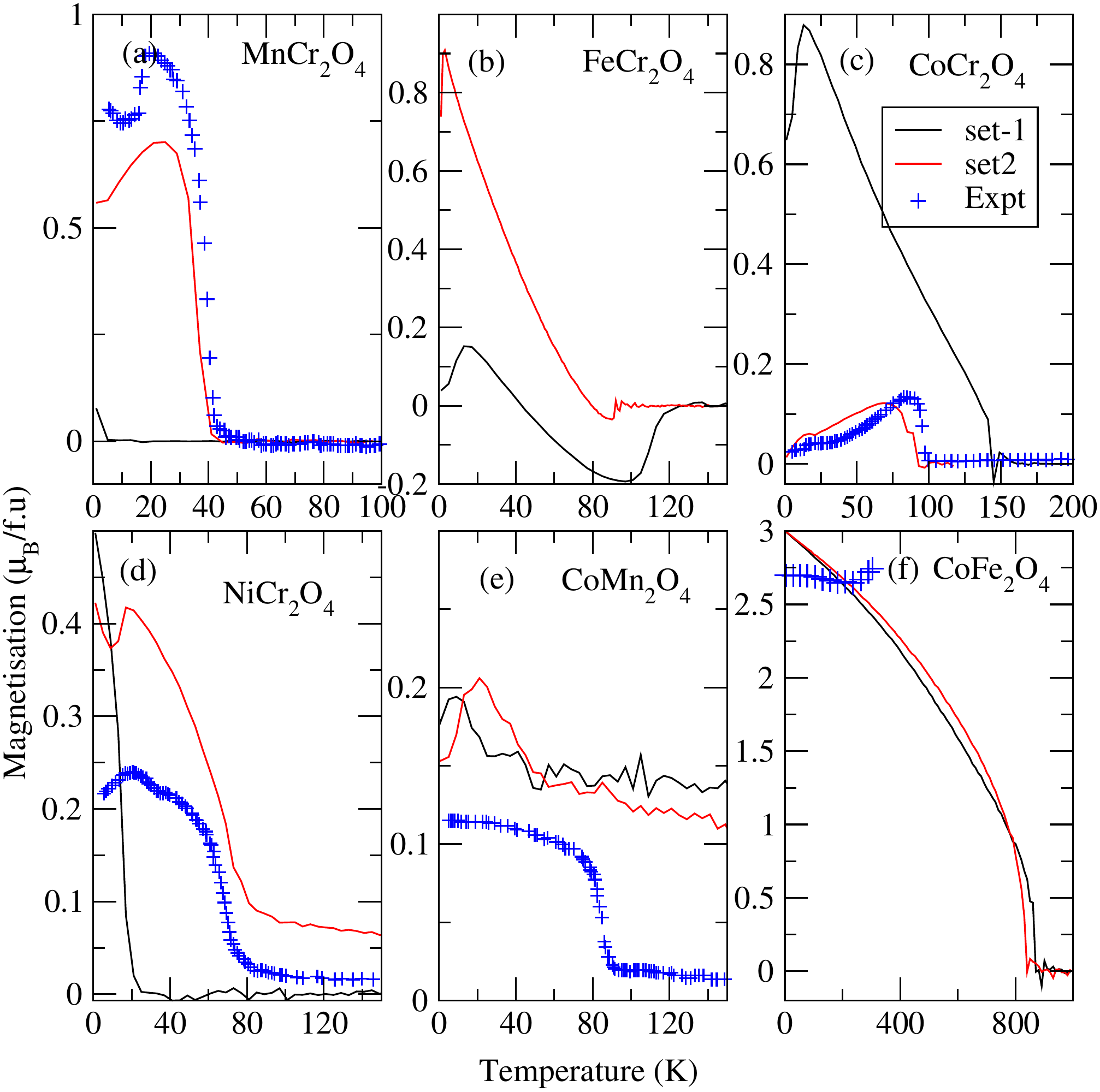}
\caption{Temperature dependence of magnetization for six spinel compounds. Black (Red) line shows the calculated magnetization using set-1 (set-2) interaction parameters. Plus symbol indicates experimental data.\cite{ACr2O4-pol,{CoMn2O4-exp}}  }
\label{fig1}
\end{figure}
\end{center}

 Figure \ref{fig1} shows a comparison of theoretical and experimental temperature dependence of magnetization for six compounds. The black (red) line indicates the calculated magnetization using set-1 (set-2) exchange parameters. Solid plus symbols show the experimental data, wherever available. It is to be noted that, for our prime compound $CoCr_2O_4$, the calculated magnetization using set-2 exchange interactions matches fairly well with those of experimental data.\cite{padam-Fe} Comparing the set-1 and set-2 parameters in this case, we found that $Cr$-$Cr$ interactions are relatively stronger in set-2 than set-1, while $Co$-$Cr$ and $Co$-$Co$ interactions in set-2 are relatively weaker. In fact, $Co$-$Co$ pairs are hardly interacting in the set-2. Table I also display the stability parameter ($u$) for all the six compounds. $u$ turn out to be 1.23 (2.00) using set-1 (set-2) exchange parameters for $CoCr_2O_4$. In case of $MnCr_2O_4$, all the interactions in set-2  are weaker compared to those in set-1. M vs T data calculated using set-1 parameters, in this case, is grossly  off as compared to experimental data. The value of $u$ calculated using set-1 (set-2) parameters is 1.81 (1.5). Both these values lie beyond the stability range (0.88$<$ $u$ $<$ 1.3 ). Interestingly the average $<u>$  calculated by Tomiyasu\cite{Tomiyasu} using the neutron scattering data, within the generalized Luttinger-Tisza \cite{GLT} method, for $CoCr_2O_4$ and $MnCr_2O_4$ are 2.00 and 1.50 which matches exactly with our calculated $u$-values. In case of $NiCr_2O_4$, the simulated magnetization which best matches with  experimental values require negligibly small $\hat{J}_{AA}$ interactions, as in the previous two cases. We don't have any experimental magnetization data for $FeCr_2O_4$. Interestingly from Figure \ref{fig1}(b), the magnetization curve calculated from set-1 shows a magnetic compensation at around $T_{comp}$= 40 K and magnetization changes its sign at this temperature. As we do not have any experimental evidence for such magnetic compensation for pure $FeCr_2O_4$ compound, therefore we calculated another set of interaction parameters (set-2), which does not show such compensation. In set-1, the value of $\hat{J}_{BB}$ and $\hat{J}_{AB}$ are close ($\hat{J}_{BB}$ is slightly higher than $\hat{J}_{AB}$). One way to remove the magnetic compensation effect is to choose $\hat{J}_{AB}$ $>$ $\hat{J}_{BB}$ which is what we have chosen in set-2. The calculated $u$ parameter  for $FeCr_2O_4$ and $NiCr_2O_4$ using set-1 are 1.35 and 1.81 which become 0.95 and 2.10 when set-2 parameters are used. Using set-2 parameters, the calculated $u$ value  is found to lie within the stability range, while for $NiCr_2O_4$, $u$ is far beyond the stability. The calculated  magnetic transition temperature ($T_c$) is also tabulated in Table \ref{table1} along with the experimental values. It is to be noted that, $T_c$ for $MnCr_2O_4$ is calculated to be 40 K(42 K) using set-1(set-2) exchange parameters whereas the magnetization of different sub-lattice cancel each other out and compensates the net moments for temperature above 4 K. At very low temperatures, it shows some finite moments. Similar behavior has also been observed in case of $FeCr_2O_4$, where the transition occurs at 103 K but just above 93 K total magnetization drops to zero.   
  
\begin{table*}
    \caption{\label{table2} For the six systems, calculated inclination angles ($\theta_A$, $\theta_{B_1}$, $\theta_{B_2}$) of the ground state magnetic order, type of spin order, polarization and conical transition temperature ($T_s$). Experimental data are given, wherever available.\cite{Tomiyasu,CoMn2O4-ang,CoFe2O4-spin}}  
\begin{center}
        \begin{tabular}{|c|c|c|c|c|c|c|c|} \hline \hline
                System    &   &\multicolumn{3}{c|}{Average inclination angle} &  Type of  &  Polarisation  & $T_s$ \\ 
                          &   &$\theta_{A}$ &$\theta_{B_1}$    &$\theta_{B_2}$&  spin order &      &  (K)   \\ \hline                        
                          &   & (Degree) & (Degree) & (Degree) &    &  ($\frac{\mu C}{m^2}$)     &   \\ \hline \hline 
                $MnCr_2O_4$& set 2& 132     & 85   & 77    &  Conical    & 4.9  & 4\\ 
                           & Expt.   & 152     & 95   & 11    &  Conical    & -    & 16\\ \hline
                $FeCr_2O_4$& set 2& 164        & 14   & 16     &  Conical    & 3.3  & 0\\  \hline 
                $CoCr_2O_4$& set 2& 142        & 83   & 40     &  Conical    & 1.8  & 16\\ 
                           & Expt.& 132        & 109  & 28     &  Conical    &  -   & 24\\ \hline
                $NiCr_2O_4$& set 2& 144        & 84   & 37     &  Conical    & 0.9  & 17    \\ \hline     
                $CoMn_2O_4$&set 2 & 90         & 141  & 38     & $A$ is the resultant of $B_1$ and $B_2$      &   0.1  & 0 \\ 
                          &Expt.  & 90         & 151  & 61     & $A$ is the resultant of $B_1$ and $B_2$               &        & 0 \\ \hline
                $CoFe_2O_4$&set 2 & 179        & 1    & 1      & $A$ is anti parallel to $B_1$ and $B_2$      &   0.0  & 0 \\ 
                          &Expt.  & 180        & 0    & 0      & $A$ is anti parallel to $B_1$ and $B_2$      &        & 0  \\ \hline \hline
            \end{tabular}
        \end{center}
\end{table*}

 \subsection{Magnetic order} 
Table \ref{table2} shows the calculated cone angle, types of spin order, polarisation and transition temperature ($T_s$)
 for the six systems. These properties are calculated using set-2 interaction parameters. Experimental data are shown wherever available. There are three cone angles $\theta_{A}$, $\theta_{B_1}$ and $\theta_{B_2}$ based on sites $A$, $B_1$, and $B_2$ respectively. Notably, the simulated value of the cone angles matches fairly well with those of experiments.\cite{Tomiyasu} Four systems $ACr_2O_4$ ($A$= $Mn$, $Fe$, $Co$ and $Ni$) show conical spin order, as also observed experimentally. For $CoMn_2O_4$, vector corresponding to $\theta_{A}$ is the resultant of those for $\theta_{B_1}$ and $\theta_{B_2}$. In case of $CoFe_2O_4$, however vector for  $\theta_{A}$ is antiparallel to those of $\theta_{B_1}$ and $\theta_{B_2}$. These magnetic orderings are in fair agreement with the experimental observation.\cite{CoMn2O4-ang}

\begin{center}
\begin{figure}[b]
\includegraphics[scale=0.4]{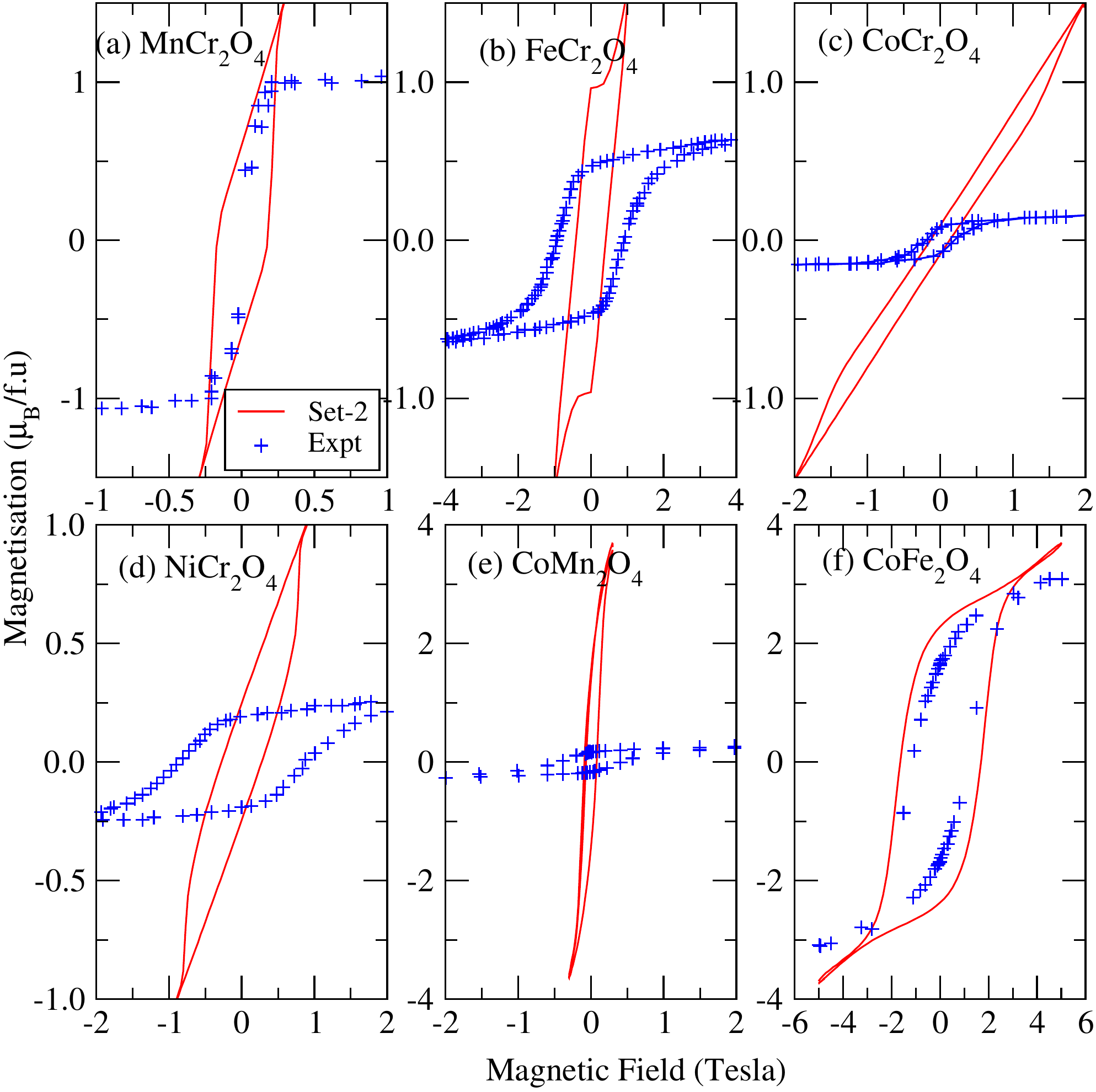}
\caption{For the six compounds, simulated and experimental\cite{ACr2O4-pol,CoMn2O4-exp,New-Exp} hysteresis curve. The simulated data is calculated with set-2 interactions parameters. }
\label{fig2}
\end{figure}
\end{center}

 \begin{center}
\begin{figure*}[t]
\includegraphics[scale=0.25 ]{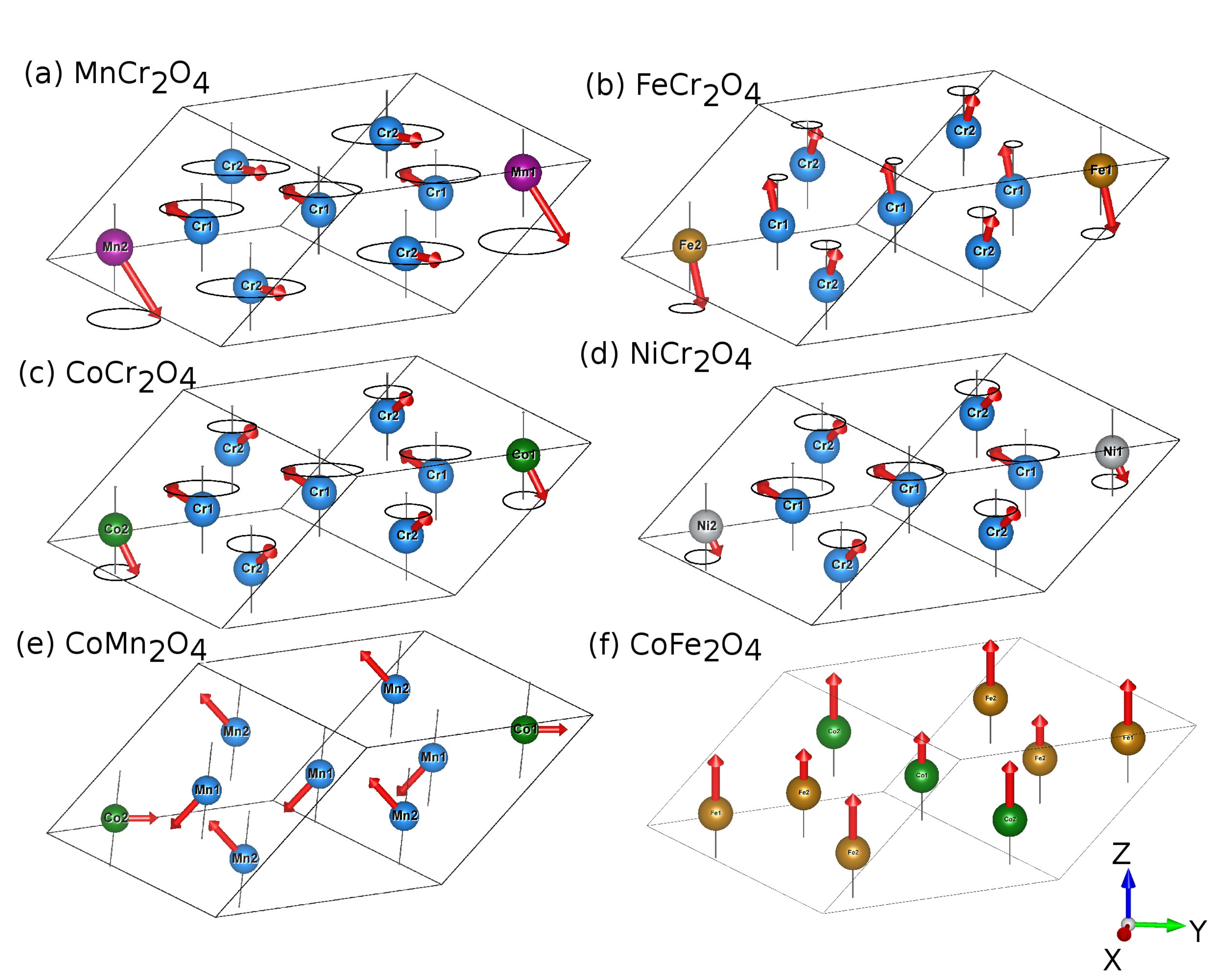}
\caption{The calculated magnetic order for six spinel compounds (a) $MnCr_2O_4$ (b) $FeCr_2O_4$ (c) $CoCr_2O_4$ (d) $NiCr_2O_4$ (e) $CoMn_2O_4$ (f) $CoFe_2O_4$.}
\label{Fig3}
\end{figure*}
\end{center}   

For $FeCr_2O_4$, the $\hat{J}_{BB}/\hat{J}_{AB}$ is nearly 1.02 for set-1 which reduce to 0.71 for set-2. This decreases the geometrical frustration and therefore the cone angle at $B$-site decreases. This in turn increase the magnetization along the positive z-direction. This also helps to uplift the magnetization curve and removes the magnetic compensation.  It is to be noted that the calculated polarisation ($\overrightarrow{P}$) and $T_s$ falls in the reasonable range.

  \subsection{Compounds having no conical order: $CoMn_2O_4$ and $CoFe_2O_4$} 
From Table \ref{table1}, the first principles calculated exchange interaction in $CoMn_2O_4$ has a strong anisotropy because it crystallizes in a tetragonal structure whereas all the other compounds are cubic. Due to stretching along the z-direction and compression in the $xy$ plane, $\hat{J}_{BB}$ in the $xy$ plane becomes much stronger and those out of a plane turn weaker. In Table 1, (I) and (O) refers to in-plane and out of plane interaction respectively. Therefore at very low temperatures, all the spins lie in the $xy$ plane and as temperature crosses $T_c$, they get completely randomized. In Figure \ref{fig1}, the calculated magnetization is plotted along with the experimental curve. For $CoFe_2O_4$, the ground state is collinear which corroborates with the fact that  $\hat{J}_{AB}$ is much stronger than $\hat{J}_{BB}$.  Interestingly, because this compound crystallizes in inverse spinel structure, which is not the case for the other five compounds, $Fe$ sits at both $A$-site and $B$-site with antiparallel alignment. This cancels out the magnetization from $Fe$ and the observed magnetization is mostly due to the magnetic moments of collinear $Co$ spins.  
Figure \ref{Fig3} shows a pictorial diagram of the calculated magnetic spin orders for all the six spinel compounds.

 \subsection{Hysteresis} 
Figure \ref{fig2} shows the calculated hysteresis (red line) for all six compounds using the set-2 interaction parameters. Experimental data are shown by plus symbol (blue). It is clear that for the compounds $ACr_2O_4$, the experimental curves reach the saturation magnetization at a relatively smaller magnetization value as compared to the calculated ones. This may be due to the conical spin spiral developed in these four compounds which reduce their magnetization. Another reason can be the neglect of higher neighbor interactions in our Monte-Carlo simulation, which probably are not small enough and can affect the more sensitive results such as the hysteresis curve. In case of $CoFe_2O_4$, hysteresis curve is quite sensitive to the interaction parameters used, while magnetization curve hardly changes. Figure \ref{fig2}(f) shows the hysteresis curves calculated from the set-2 interaction parameters which matche fairly well with experiment.    In contrast, both our calculated  M Vs. T and hysteresis for $CoMn_2O4$ are somewhat different compared to experiment. This may be due to the fact that, in experimental sample of $CoMn_2O4$,\cite{CoMn2O4-exp}   21 \% of $Co$ atoms are observed to interchange its positions with $Mn$. Such swapping is not considered in our calculations. 
 
\begin{center}
\begin{figure}[t]
\includegraphics[scale=0.4]{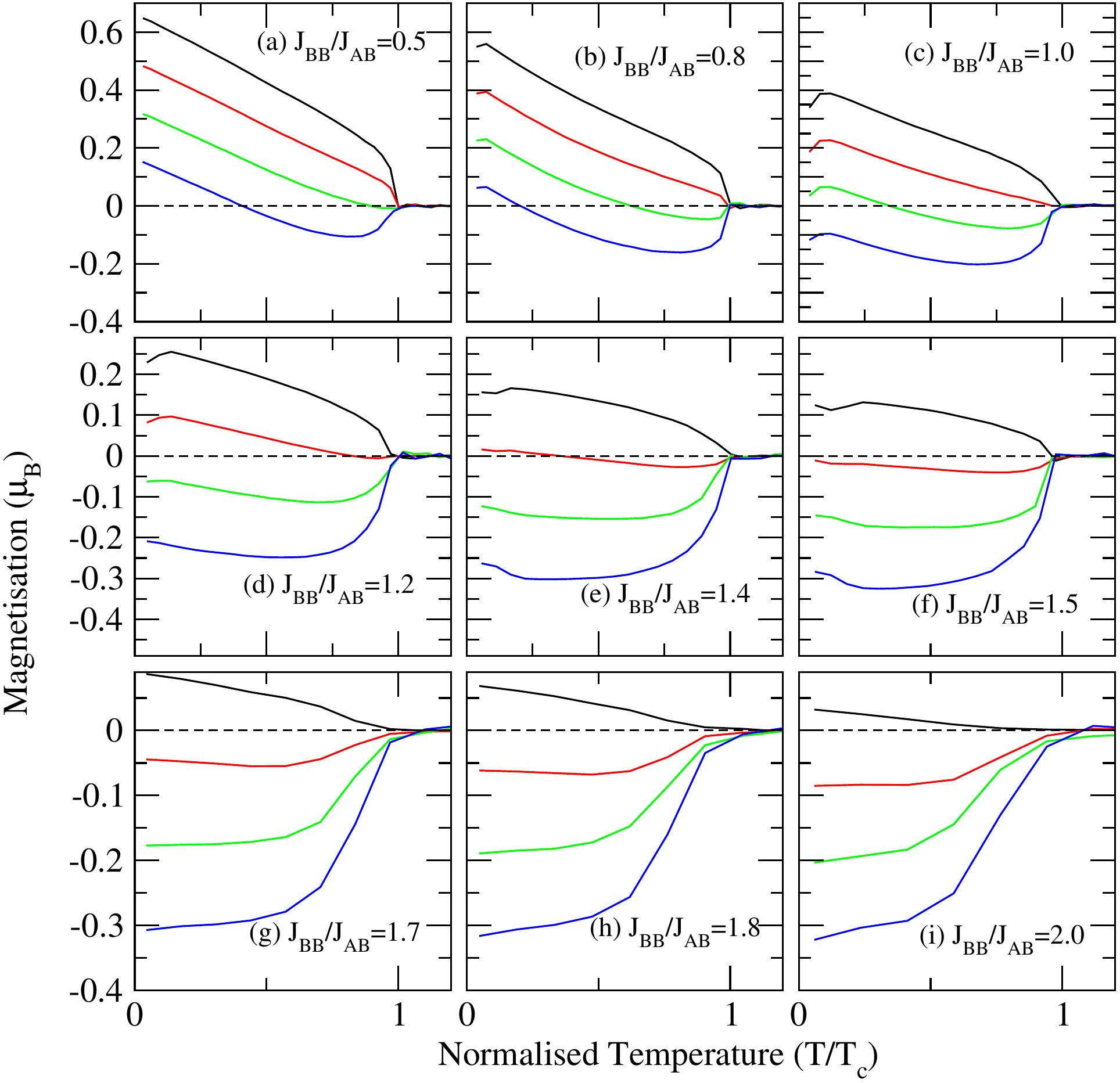}
\caption{Total magnetisation vs. $T$/$T_c$  at various values of ($\hat{J}_{BB}$/$\hat{J}_{AB}$)  for $MnCr_2O_4$ (blue), $FeCr_2O_4$ (green), $CoCr_2O_4$ (red) and $NiCr_2O_4$ (black).}
\label{Fig4}
\end{figure}
\end{center}   

    \subsection{Polarisation}
    Polarisation($\overrightarrow{P}$) for $ACr_2O_4$ is calculated using Eq (2). The proportionality constant `$a$' is taken to be 0.03 $\frac{\mu C}{m^2}$. $\overrightarrow{P}$ is calculated using set-2 exchange parameters, which involve $BB$, $AB$ and $AA$ type of 1st neighbour interactions. Yao \textit{et al.}\cite{Yao-2009,Yao-2009-2,Yao-2010,Yao-2011,Yao-2013,Yao-2017} have also reported the calculation of $\overrightarrow{P}$ using only $BB$-type neighbour interaction. We observed that, inclusion of $AB$ and $AA$
 (in addition to $BB$) interactions help to achieve the stable conical spin spiral order easily. Singh \textit{et al.}  measured the polarisation for both $CoCr_2O_4$ and $FeCr_2O_4$,\cite{FeCr2O4-pol} and found the magnitude of $\overrightarrow{P}$ for $FeCr_2O_4$ to be 10-12 times larger. This indicates that the choice of `$a$' value is crucial in the theoretical simulation of $\overrightarrow{P}$. As we do not have much information for the rest of the compounds, for simplicity we have taken `$a$' to be  0.03 $\frac{\mu C}{m^2}$ for all the compounds in the calculation of $\overrightarrow{P}$. It is to be noted that, as the magnitude of $A$-site spin decreases, the polarisation  also decreases. In $CoFe_2O_4$, the calculated polarization is nearly zero as all the spins are collinear.  For the compound $CoMn_2O_4$, the simulated polarisation is found to be quite small in magnitude,  ~0.1 $\frac{\mu C}{m^2}$. The critical temperature $T_s$ below which the polarisation can be measured  are also listed in Table \ref{table2}. In all the 4 compounds, except $CoCr_2O_4$, $T_s$, the value calculated using set-1 exchange parameters is higher than the set-2 parameters. This suggests that set-2 is giving more accurate cone-angle. It is important to note that, even the set-2 parameters are only a set of effective interaction parameters where higher-order interactions can be considered to be included within a mean-field scheme. This may be one of the reasons for some discrepancies.

  \subsection{Magnetic compensation}
It has been observed that some ferrimagnets have a certain critical temperature, below the ferri-para transition region ($T_c$), called the magnetic compensation temperature  ($T_{comp}$),  where the magnetization curve crosses the zero temperature axes. At $T=T_{comp}$, the antiferromagnetic spins of different sublattices just cancel each other out to give a compensating net zero magnetization. The magnetization just below and above $T_{comp}$ have opposite signs. 

The compensation has not been reported in any of a pristine spinel compounds $MnCr_2O_4$, $CoCr_2O_4$ and $NiCr_2O_4$ but is detected in some of their substituted counterpart. It is not easy to simulate the substituted systems, as we need to evaluate a new set of exchange parameters between the substituting magnetic atom and the rest of the atoms of the pristine compound. Also, the final result sensitively depends upon the substituting sites chosen in the Monte Carlo simulation. We chose to address this problem in the future. However, to check the possibility of magnetic compensation, we have calculated the magnetization vs. T for various interaction strengths $\hat{J}_{BB}$/$\hat{J}_{AB}$ from 0.5 to 2.0. This is shown in Fig. \ref{Fig4}  for the four compounds $ACr_2O_4$. These parameters can be thought of as effective interactions when the pristine compounds are substituted with a foreign element.

For $CoCr_2O_4$ (red curve in Fig. \ref{Fig4}), there is a clear indication of magnetic compensation temperature of $T$/$T_c$ = 0.3 for $\hat{J}_{BB}$/$\hat{J}_{AB}$= 1.4. Any interaction with $\hat{J}_{BB}$/$\hat{J}_{AB}$$>$1.4, makes the system non-compensating. For $\hat{J}_{BB}$/$\hat{J}_{AB}$$<$1.4, $T_{comp}$ increases towards higher T-side and again become non-compensating for $\hat{J}_{BB}$/$\hat{J}_{AB}$$<$1.0. Similar trend is found for $MnCr_2O_4$ and $FeCr_2O_4$ as well, but with different $T_{comp}$.  For $NiCr_2O_4$, we could not find any compensation temperature between the range 0.5 $\leq$  $\hat{J}_{BB}$/$\hat{J}_{AB}$ $\leq$ 2.0.

\subsection{Origin Magnetic compensation}
\begin{center}
\begin{figure}[t]
\includegraphics[scale=0.4]{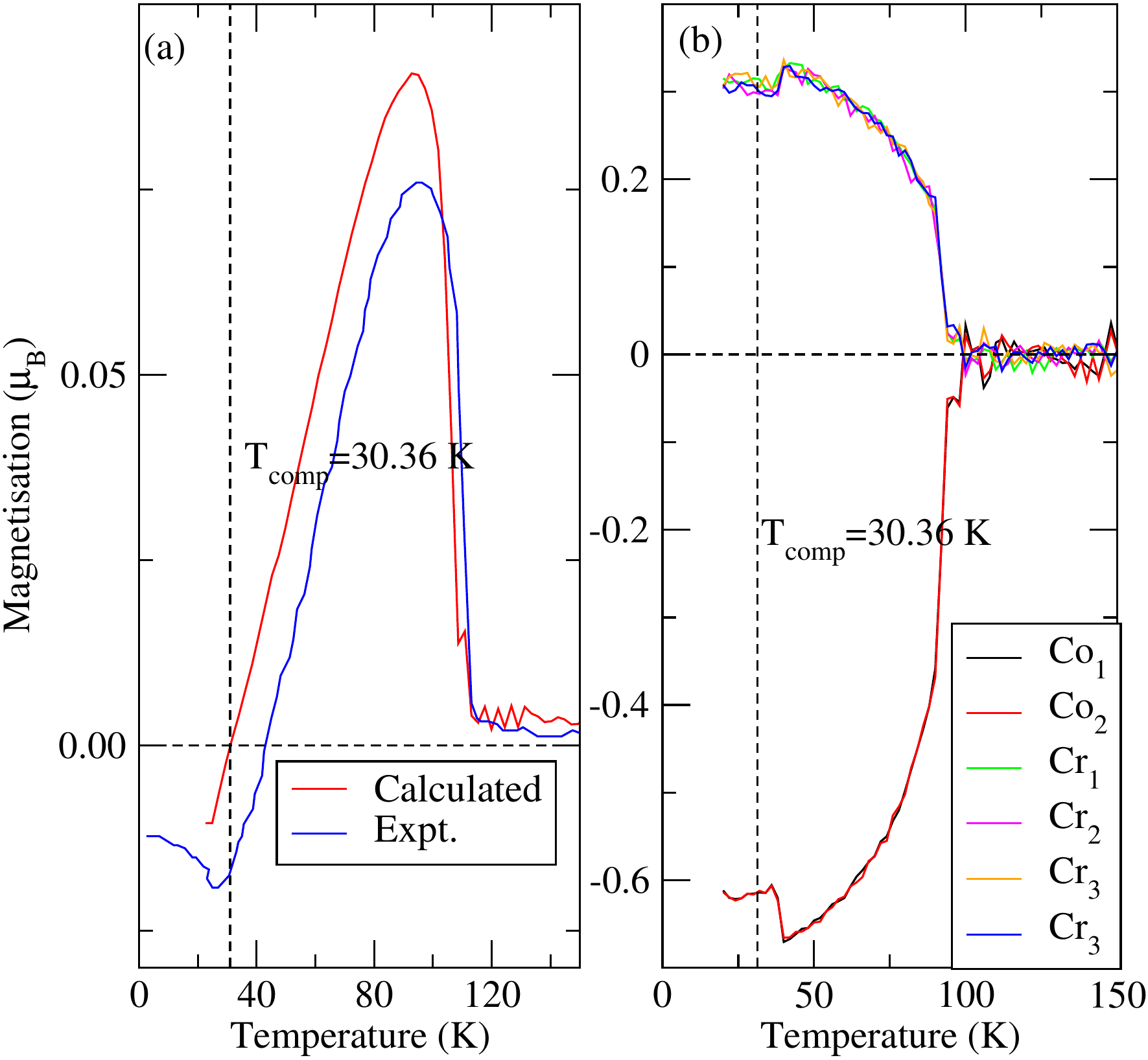}
\caption{(Left) Calculated M vs. T curve (red) for $CoCr_2O_4$ with effective interaction parameters $\hat{J}_{BB}$/$\hat{J}_{AB}$=1.41, along with the experimental curve (blue) for $Co(Cr_{0.95}Fe_{0.5})_2O_4$. (Right) Calculated atom-projected M vs. T curve for $CoCr_2O_4$.}
\label{Fig5}
\end{figure}
\end{center}

The origin of magnetic compensation lies in the cancellation of magnetization between A- and B-sites which, in turn, depends on the exchange interactions. In Fig. \ref{Fig5}, the total and atom projected magnetizations  (for $\hat{J}_{BB}$/$\hat{J}_{AB}$=1.41) are  plotted in the left and the right panels respectively, for $CoCr_2O_4$. This indicates that one can dictate the variation in $T_{comp}$ by tuning the magnetization at different sublattices. Substitution/doping is a unique way to modify the magnetization of a given system. This can affect the magnetization in two different ways: (i) the substituted magnetic atom manipulate the magnetization of that sublattice (ii) the exchange interaction between the substituted atoms with the rest of the atoms changes the spin alignment and hence the magnetization.

By mimicking the substituting effect via an effective change in the exchange interactions, we found that as we increase $\hat{J}_{BB}$, the frustration in the $B$-sublattice increases and the magnetic spins of $Cr$-atoms start to deviate from the collinear state.  This reduces the magnetization from the $B$ sublattice. As a result, the total magnetization increases. Since going from $MnCr_2O_4$ to $NiCr_2O_4$, the $A$ site magnetization reduces,  the total magnetization increases in the direction of the magnetic orientation of $B$ sublattices. Therefore, to get the compensation temperature in $NiCr_2O_4$, we need to increase the $\hat{J}_{BB}$ interaction which creates more frustration in the B sublattice reducing its magnetization. This, in turn, will help the total magnetization to cross the temperature axis at some point. 

\subsection{Exchange Bias in $CoCr_2O_4$} 
\begin{center}
\begin{figure}[t]
\includegraphics[scale=0.4]{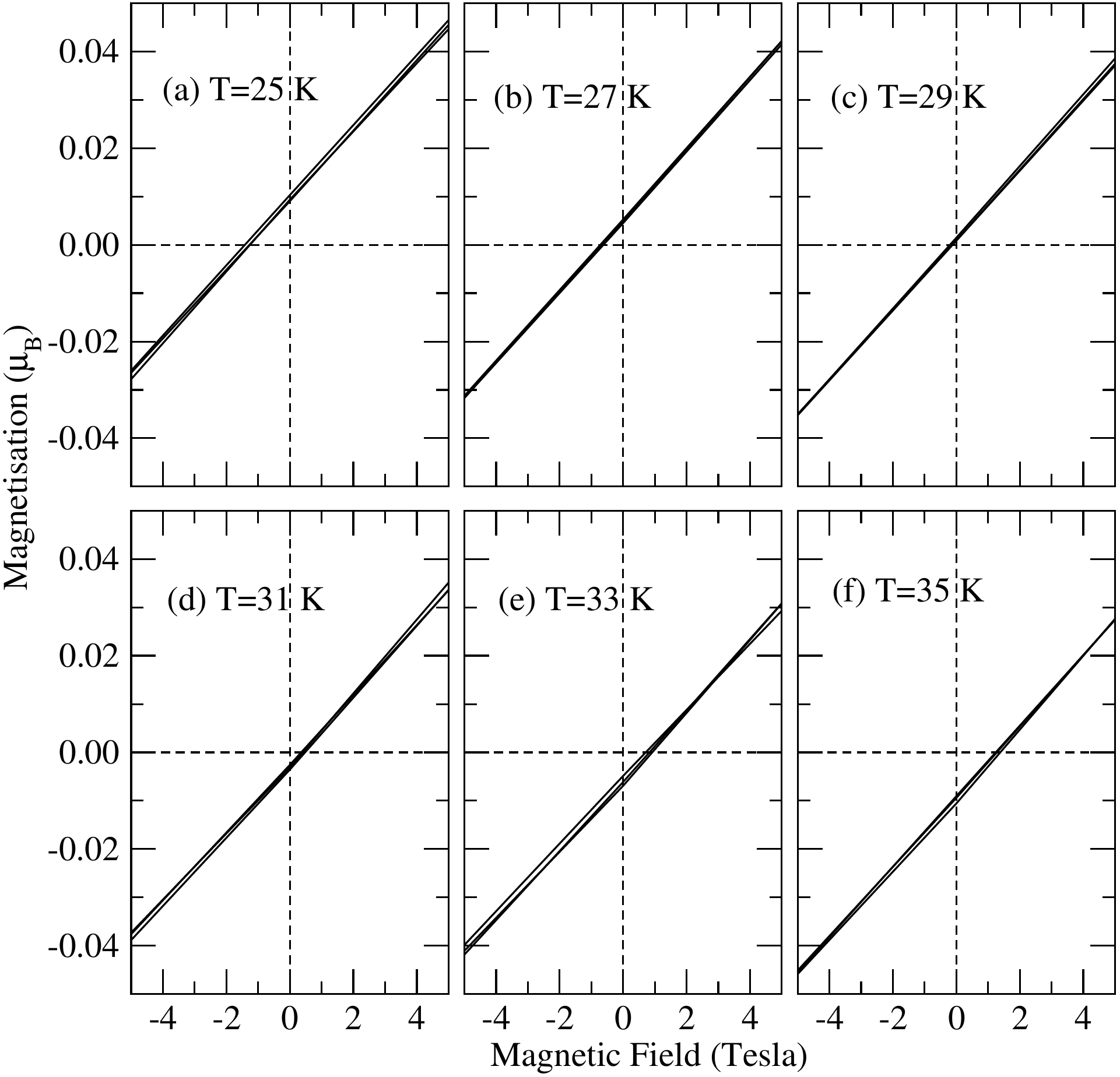}
\caption{ Sign reversible exchange bias effect (shift of the origin of hysteresis with varing temperature) in $CoCr_2O_4$ with $\hat{J}_{BB}$/$\hat{J}_{AB}$ =1.41.}
\label{Fig6}
\end{figure}
\end{center}

Exchange bias is a phenomenon that shifts the origin of hysteresis on the magnetic axis. For most of the memory device and the device based on spintronics application need a layer having exchange bias so that it fixes the magnetic state with surrounding magnetic fluctuation. It has been reported that very close to $T_{comp}$, exchange bias is observed in the $Fe$ substituted $CoCr_2O_4$.\cite{padam-Fe} With a similar motivation as before, we have studied the appearance of exchange bias by mimicking the effect of substitution via the change in effective interactions. Figure \ref{Fig6} shows the shift in the hysteresis as a function of varying temperature with $\hat{J}_{BB}$/$\hat{J}_{AB}$=1.41 ($\hat{J}_{BB}$=-4.00, $\hat{J}_{BB}$=-2.83). These parameters can only be taken in an average sense representing the mean-field estimate of the exchange interactions for $Fe$-substituted $CoCr_2O_4$. Interestingly, at around 30.36 K, sign reversible exchange bias is observed. The transition temperature agrees fairly well with the magnetic compensation temperature, as observed experimentally.\cite{{padam-EB}}  Experimentally a magneto-structural correlation
 has been observed at around $T_{comp}$\cite{ram-Mn,padam-MS-corr1,debashish-ram}. As  we have not considered the magneto-structural correlation in our calculation but we successfully able to detect exchange bias effect. Therefore we conclude that the exchange bias created in these substituted compounds is purely due to the magnetic spin order developed at low temperature and is independent of magneto-structural correlations.

\section{Conclusion}
In summary, we have investigated the possibility of conical magnetic order in a series of six $AB_2O_4$ spinel compounds using Monte-Carlo simulation. These calculations are done with a careful choice of two sets of interaction parameters: (i) parameter set-1 obtained from self-consistent first principles-based DFT simulation and (ii) parameter set-2, which closely reproduce the experimental magnetization. set-2 parameters are further used to evaluate the rest of the magnetic properties such as hysteresis, magnetic order, exchange bias, etc. Considering $CoCr_2O_4$ as a representative system, we have been able to reproduce the correct angle of conical order and the stability parameter $u$, as observed. The estimated polarisation and the transition temperature agree fairly well with the experiment. The effect of Fe substitution in $CoCr_2O_4$ is simulated by mimicking a different set of exchange interactions. These parameters can be considered as the effective interactions, within a mean-field sense, representing the Fe substituted system $Co(Cr_{0.95}Fe_{0.05})_2O_4$. We found that this compound indeed shows a sign reversible exchange bias effect at around $T_{comp}$  =30.4 K, as observed experimentally, which is purely magnetic origin as we have not considered magneto-structural correlations observed around $T_{comp}$ in experiment but successfully able to mimic exchange bias phenomena. We have also simulated $CoMn_2O_4$ and $CoFe_2O_4$, and found no conical magnetic order and polarisation, as observed. The spin-current model which is used in our calculation works quite well for very low magnetic field and therefore with high magnetic field, the magnetisation will not saturate as observed in experiment. Similarly, this model is not thermally stable and the polarisation drops quite fast compare to experiments. Therefore a better model is needed to work in high magnetic field and high temperature. However, this model shows its potential by getting the nearly similar cone angle of the atomic spins as in experiments and also able to mimic exchange bias phenomena and magnetic compensation quite well.              

\section{Acknowledgments}
We thank IIT Bombay for lab and computing facilities. AA acknowledge IRCC early carrier research award project, IIT Bombay (RI/0217-10001338-001) to support this research. DD acknowledge  financial support provided by the Science and Engineering Research Board (SERB) under the National Post Doctoral Fellowship, sanction order number PDF/2017/002160.

\end{document}